\documentclass[a4paper,fleqn,usenatbib]{mnras}

\usepackage[T1]{fontenc}
\usepackage{ae,aecompl}

\usepackage{graphicx}	
\usepackage{amsmath}	
\usepackage{amssymb}	
\usepackage{times}
\usepackage{graphics, subfigure}
\usepackage{epsfig}
\usepackage{multirow}
\usepackage{ulem}
\usepackage{pdfpages}

\def\simlt{\mathrel{\hbox{\rlap{\hbox{\lower4pt\hbox{$\sim$}}}\hbox{$<$}}}}
\def\simgt{\mathrel{\hbox{\rlap{\hbox{\lower4pt\hbox{$\sim$}}}\hbox{$>$}}}}

\newcommand{\mysim}{\mathord{\sim}}

\newcommand{\myapprox}{\mathord{\approx}}
\newcommand{\nimass}{M_\text{Ni56}}
\newcommand{\MWD}{M_{\rm{WD}}}

\title[low-luminosity double detonation]{Confronting double-detonation sub-Chandrasekhar models with the low-luminosity suppression of Type Ia supernovae}

\author[A. Ghosh and D. Kushnir]{
Arka Ghosh$^{1}$\thanks{E-mail: arka@post.bgu.ac.il} and Doron Kushnir$^{2}$
\\
$^{1}$ Physics Department, Ben-Gurion University of the Negev, Be'er-Sheva 84105, Israel \\
$^{2}$Dept. of Particle Phys. \& Astrophys., Weizmann Institute of Science, Rehovot 76100, Israel\\
}

\date{Accepted XXX. Received YYY; in original form ZZZ}

\pubyear{2019}

\begin{document}
\label{firstpage}
\pagerange{\pageref{firstpage}--\pageref{lastpage}}
\maketitle

\begin{abstract}
Type Ia supernovae (SNe Ia) are likely the thermonuclear explosions of carbon-oxygen (CO) white-dwarf (WD) stars, but their progenitor systems remain elusive. Recently, Sharon \& Kushnir (2022) used The Zwicky Transient Facility Bright Transient Survey to construct a synthesized $^{56}$Ni mass, $\nimass$, distribution of SNe Ia. They found that the rate of low-luminosity ($\nimass\approx0.15\,M_{\odot}$) SNe Ia is lower by a factor of $\mysim10$ than the more common $\nimass\approx0.7\,M_{\odot}$ events. We here show that in order for the double-detonation model (DDM, in which a propagating thermonuclear detonation wave, TNDW, within a thin helium shell surrounding a sub-Chandrasekhar mass CO core triggers a TNDW within the core) to explain this low-luminosity suppression, the probability of a low-mass ($\approx0.85\,M_{\odot}$) WD explosion should be $\mysim100$-fold lower than that of a high-mass ($\approx1.05\,M_{\odot}$) WD. One possible explanation is that the ignition of low-mass CO cores is somehow suppressed. We use accurate one-dimensional numerical simulations to show that if a TNDW is able to propagate within the helium shell, then the ignition within the CO core is guaranteed (resolved here for the first time in a full-star simulation), even for $0.7\,M_{\odot}$ WDs, providing no natural explanation for the low-luminosity suppression. DDM could explain the low-luminosity suppression if the mass distribution of primary WDs in close binaries is dramatically different from the field distribution; if the Helium shell ignition probability is suppressed for low-mass WDs; or if multidimensional perturbations significantly change our results.
\end{abstract}




\section{Introduction}
\label{sec:Introduction}

\begin{figure}
\includegraphics[width=0.48\textwidth]{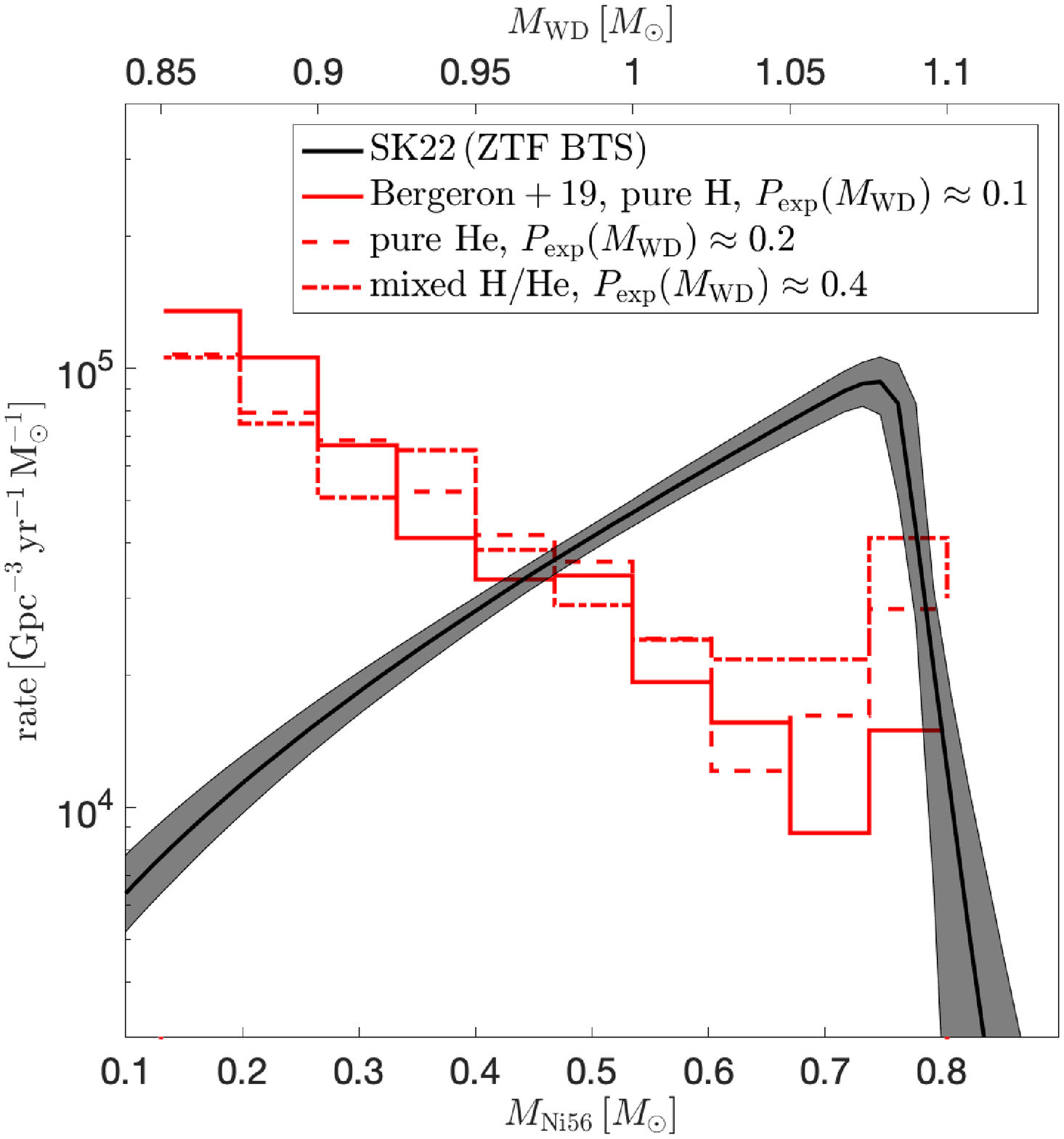}
\caption{The $\nimass$ distribution of SNe Ia \citep[black,][]{Sharon2022} as constructed from the ZTF BTS \citep[][]{Fremling2020,Perley2020}. Shown are the 68-per-cent confidence level of the distribution (shaded area) and its median (black line). The rate of low-luminosity ($\nimass\approx0.15\,M_{\odot}$) SNe Ia is lower by a factor of $\mysim10$ compared to the more common $\nimass\approx0.7\,M_{\odot}$ events. The upper $x$-axis is transformed with $\nimass/M_{\odot}=2.693(\MWD/M_{\odot})-2.165$, derived from a fit to the SCD solar-metallicity results of \citet{Kushnir2020}. Red lines: The predicted $\nimass$ distribution in DDM given the mass distribution of WDs within $100\,\rm{pc}$ from the Sun \citep[][]{Bergeron2019}, as constructed from \textit{Gaia} data release 2 \citep[][]{Gaia2018}. The WDs have been fitted under the assumption of a pure hydrogen envelope (solid), a pure helium envelope (dashed), or a mixed hydrogen/helium envelope (dot-dashed). Since the relation between $\MWD$ and $\nimass$ is linear for DDM in the presented range, the predicted SNe Ia rate in DDM is just the WD's mass distribution, up to a scaling (assuming the chance that a WD explodes, $P_{\rm{exp}}$, depends weakly on its mass in the presented range). We normalize the predicted rate with $P_{\rm{exp}}\approx0.1(0.2,0.4)$, which leads to a similar total SNe Ia rate \citep[since $\approx10(5,2.5)\%$ of WDs are within the mass range $0.85-1.1\,M_{\odot}$ for the pure hydrogen (pure helium, mixed composition) assumption,][and $\approx1\%$ of WDs should explode to match the SNe Ia rate]{Bergeron2019}. Low-mass WDs ($\approx0.85\,M_{\odot}$ WDs, the progenitors of $\nimass\approx0.15\,M_{\odot}$ in the DDM) are $\mysim10$-fold more common than high-mass WDs ($\approx1.05\,M_{\odot}$ WDs, the progenitors of $\nimass\approx0.7\,M_{\odot}$ in the DDM), where the ratio depends somewhat on the envelope composition assumption. In order for the DDM to agree with the low-luminosity suppression, the probability of low-mass WDs exploding should be much lower (by a factor of $\mysim100$) than that of high-mass WDs. Note that a very high efficiency, $P_{\rm{exp}}(\MWD\approx1.1\,M_{\odot})\sim1$, of the DDM in exploding $\MWD\approx1.1\,M_{\odot}$ WDs is required in order to agree with the observed rate of $\nimass\approx0.7\,M_{\odot}$.
\label{fig:rate}}
\end{figure}

Type Ia supernovae (SNe Ia) are likely the thermonuclear explosions of carbon-oxygen (CO) white-dwarf (WD) stars, but their progenitor systems remain elusive \citep[see][for a review]{Maoz2014}. Recently, \citet{Sharon2022} used The Zwicky Transient Facility Bright Transient Survey \citep[ZTF BTS;][]{Fremling2020,Perley2020} to construct the luminosity function and the synthesized $^{56}$Ni mass, $\nimass$, distributions of SNe Ia. They found that the distributions are unimodal, with their peaks in agreement with previous results, but feature a much lower rate of dim and luminous events (see black line in Figure~\ref{fig:rate}). We focus in this paper on the $\mysim10$-fold low-luminosity ($\nimass\approx0.15\,M_{\odot}$) SNe Ia suppression rate compared to the more common $\nimass\approx0.7\,M_{\odot}$ events. This low-luminosity suppression can be used to discriminate between SNe Ia models. Specifically, we show that the prediction of the "double-detonation model" \citep[DDM;][]{Nomoto1982a,Nomoto1982b,Livne1990,Woosley1994} is in apparent contradiction with the low-luminosity suppression.

\vspace{10mm}

The DDM  has been studied extensively \citep{Hoeflich1996,Nugent1997,Bildsten2007,Fink2007,Fink2010,Kromer2010,Woosley2011,Moore2013,Shen2014,ShenMoore2014,Polin2019,Townsley2019,Gronow2020,Boos2021,Gronow2021a,Gronow2021b}. In this model, a thermonuclear detonation wave (TNDW) propagates within a helium shell that surrounds a sub-Chandrasekhar mass CO core. The outer TNDW launches an imploding shock wave that accelerates as it implodes and can trigger a second ignition at the CO core. The outcome of such an explosion would be very similar to a central ignition of a Sub-Chandrasekhar mass CO WD (Sub-Chandra detonation, SCD), which has been heavily studied \citep{Sim2010,Moll2014,Blondin2017,Shen2018,Bravo2019,Kushnir2020}. The $\nimass$ is controlled by the mass of the WD, $\MWD$, where $\nimass$ monotonically increases with $\MWD$ \citep[e.g.,][]{Sim2010,Moll2014,Blondin2017,Shen2018,Bravo2019,Kushnir2020}. The solar-metallicity results of \citet{Kushnir2020} can be fitted with $\nimass/M_{\odot}=2.693(\MWD/M_{\odot})-2.165$ to better than $3$ per cent accuracy in the range $0.85\le\MWD/M_{\odot}\le1.1$, which corresponds to the range $0.12\lesssim\nimass/M_{\odot}\lesssim0.8$. This transformation is used to determine the upper $x$-axis of Figure~\ref{fig:rate}. Since the relation between $\MWD$ and $\nimass$ is linear for DDM in this range, the predicted SNe Ia rate in DDM is just the WD's mass distribution, up to a scaling (assuming the chance that a WD explodes, $P_{\rm{exp}}$, depends weakly on its mass in this range).

\citet{Bergeron2019} have used WDs detected by the \textit{Gaia} mission \citep[data release 2;][]{Gaia2018} to construct the mass distribution of WDs within $100\,\rm{pc}$ from the Sun. The WDs have been fitted under the assumption of a pure hydrogen envelope, a pure helium envelope, or a mixed hydrogen/helium envelope. We normalize the results of \citet[][red lines in Figure~\ref{fig:rate}]{Bergeron2019} with $P_{\rm{exp}}\approx0.1(0.2,0.4)$, which leads to a similar total SNe Ia rate \citep[since $\approx10(5,2.5)\%$ of WDs are in the mass range $0.85-1.1\,M_{\odot}$ for the pure hydrogen (pure helium, mixed composition) assumption,][and $\approx1\%$ of WDs should explode to match the SNe Ia rate]{Bergeron2019}. As can be seen in the figure, $\approx0.85\,M_{\odot}$ WDs (hereafter \textit{low-mass WDs}, the progenitors of $\nimass\approx0.15\,M_{\odot}$ in the DDM) are more common by a factor of $\mysim10$ than $\approx1.05\,M_{\odot}$ WDs (hereafter \textit{high-mass WDs}, the progenitors of $\nimass\approx0.7\,M_{\odot}$ in the DDM), where the ratio depends somewhat on the envelope composition assumption. The distribution of $\MWD$ can be skewed towards higher masses in environments with an average lifetime much shorter than that of the solar environment. However this is probably irrelevant to SNe Ia, as half of all SNe Ia occur $>1\,\rm{Gyr}$ after their progenitor system forms \citep[see, e.g., ][and references therein]{Maoz2014}, much longer than the main-sequence lifetimes of the stars that produce a $\MWD=0.85\,M_{\odot}$. This is especially true for the range $0.15\lesssim\nimass/M_{\odot}\lesssim0.5$, which is completely dominated by old red sequence galaxies \citep{Sharon2022}. Therefore, in order for the DDM to agree with the low-luminosity suppression, the probability that a low-mass WD explodes should be much lower (by a factor of $\mysim100$) than that of a high-mass WD. Note that a very high efficiency, $P_{\rm{exp}}(\MWD\approx1.1\,M_{\odot})\sim1$, of the DDM in exploding $\MWD\approx1.1\,M_{\odot}$ WDs is required in order to agree with the observed rate of $\nimass\approx0.7\,M_{\odot}$. 

One possibility is that the probability of a WD to be involved in a binary that leads to the required conditions for the DDM to operate is much lower for low-mass WDs than for high-mass WDs. While this probability is observationally poorly constrained and its calculation is highly uncertain, the results of binary population synthesis calculations do not support this possibility. For example, \citet{Shen2017} used binary population synthesis to find at merger more systems with $\MWD\approx0.85\,M_{\odot}$ WD than systems with $\MWD\approx1.05\,M_{\odot}$ WD (see their figure 2; note that systems with a non-degenerate helium star donor are not considered in this study). Nevertheless, it is difficult to rule out the possibility that the primary WDs in close binaries have a mass distribution that is significantly different from the results of \citet{Bergeron2019}. 

Another possibility is that the ignition probability of a TNDW within the Helium shell is suppressed for low-mass WDs. While the ignition mechanism of the Helium shell is not fully understood \citep[see, e.g.,][]{Zingale2013,Jacobs2016,Glasner2018}, one could speculate that since low-mass WDs have a lower virial temperature, then a more stringent conditions are required from the progenitor binary to achieve ignition. Alternatively, high enough density for a propagating TNDW requires more massive He shells for low-mass WDS, which can also preclude some binaries.

In this paper, we study in detail the possibility that the ignition of low-mass CO cores is somehow suppressed. \citet{KLW2012} studied the connection between the converging flows and the generation of much smaller scale hot spots that may trigger a detonation wave, both for terrestrial applications and for the conditions at a typical (density of $\rho\sim10^{7}\,\textrm{g}\,\textrm{cm}^{-3}$) CO core. They estimated, without presenting detailed calculations, that the critical strength of the shock (measured by $R_9$, the radius in which the shock velocity is $10^{9}\,\textrm{cm}\,\textrm{s}^{-1}$) required for the ignition of a CO core is $R_{9}\sim0.01\,\textrm{km}$. The implication of this number is that any imploding shock that can accelerate to $\myapprox10^{9}\,\textrm{cm}\,\textrm{s}^{-1}$ at a radius that is $\gtrsim0.01\,\textrm{km}$ will ignite a TNDW in the CO core. Later on, \citet{Shen2014} verified the value $R_{9}\sim0.01\,\textrm{km}$ by using detailed numerical calculations and also studied the ignition condition for a wide range of core densities. They found that the critical $R_9$ increases with lower core densities, which may indicate that the ignition of low-mass CO cores is suppressed. We use a one-dimensional (1D) approach in Section~\ref{sec:work} to show that the work delivered by a piston, $W_p$, which mimics a TNDW in the helium shell, produces an $R_9$ that is many orders of magnitude larger than the critical $R_9$ for ignition, even for very low-mass, $\MWD=0.7-0.85\,M_{\odot}$, WDs. The large $R_9$ in these cases allows us to verify the ignition, by numerically resolving it in a full-star simulation for the first time (see discussion of previous attempts in Section~\ref{sec:discussion}). We find an upper limit of $10^{-3}\,M_{\odot}\,\textrm{MeV}\,m_p^{-1}$ to the minimal $W_p$ required for ignition, whereas the minimal helium shell that allows a propagating TNDW within it would deliver $\mysim10^{-2}\,M_{\odot}\,\textrm{MeV}\,m_p^{-1}$ \citep{ShenMoore2014}. The fact that the upper limit we provide for $W_p$ is much smaller than what could be reasonably delivered shows that the ignition is robust.  

In Section~\ref{sec:setup}, we describe the 1D setup we implemented using the hydrodynamical code VULCAN \citep[Lagrangian, hereafter V1D; for details, see][]{Livne1993IMT}, which includes a new accurate and efficient burning scheme \citep{KK2019}. In Section~\ref{sec:work}, we show that $W_p=10^{-3}\,M_{\odot}\,\textrm{MeV}\,m_p^{-1}$ is sufficient to ignite TNDW, even for very low-mass, $\MWD=0.7-0.85\,M_{\odot}$, WDs. In Section~\ref{sec:example}, we provide a 1D example of a successful core ignition with $W_p=10^{-3}\,M_{\odot}\,\textrm{MeV}\,m_p^{-1}$. The purpose of this example is to clarify the ignition process behind accelerating shock \citep{KLW2012,Kushnir2013} and to demonstrate that our numerical scheme is able to resolve this ignition. We summarise our results in Section~\ref{sec:discussion}, where we also discuss the expected multi-D modification to our 1D approach. We conclude that the DDM provides no natural explanation for the low-luminosity suppression of SNe Ia. 

In what follows, we normalize temperatures, $T_9=T\,[\,\textrm{K}]/10^{9}$, and densities, $\rho_7=\rho\,[\,\textrm{g}\,\textrm{cm}^{-3}]/10^{7}$. Some aspects of this work were calculated with a modified version of the {\sc MESA} code\footnote{Version r7624; https://sourceforge.net/projects/mesa/files/releases/} \citep{Paxton2011,Paxton2013,Paxton2015}. 
 

\section{Numerical scheme and setup}
\label{sec:setup}

In this section, we describe the 1D setup we implemented. We describe our initial setup in Section~\ref{sec:initial} and the Lagrangian numerical scheme V1D in Section~\ref{sec:V1D}. 

\subsection{Initial setup}
\label{sec:initial}

The WD profiles are constructed using a modified version of a routine by Frank Timmes\footnote{http://cococubed.asu.edu/} that includes the input physics of Appendix~\ref{sec:input}. The WDs are isothermal with an initial temperature of $T_{\textrm{WD},9}=0.01$. The initial composition is uniform throughout the WD, with mass fractions $X(^{12}\textrm{C})=X(^{16}\textrm{O})=0.4925$ and $X(^{22}\textrm{Ne})=0.015$ \citep[which corresponds to solar metallicity;][]{Kushnir2020}. The helium shell is not part of the model. 

We approximate the work delivered by the propagating TNDW within the helium shell as a 1D piston. As we explain in Section~\ref{sec:work}, our results do not depend on the exact implementation of the piston. We therefore choose a simple parametrisation for the piston velocity, which mimics the (azimuthally averaged) velocity of the interface between the helium shell and the CO core. Unless stated otherwise, the piston's initial position, $r_p$, is where the initial density is $\rho_{p,7}=0.01$, which is a typical density at the base of the helium shell. Placing $r_p$ at a lower density significantly decreases the time step and has a minimal effect on our results; see Section~\ref{sec:work}. The velocity profile is $v=-v_p(1-t/t_p)$ for $t\le t_p$ (and zero otherwise), with $v_p$ and $t_p$ free parameters. We scan a large range of $v_p\in[1,10]\times10^{3}\,\textrm{km}\,\textrm{s}^{-1}$ and we find that our results are independent of the exact chosen values.  

\subsection{The numerical scheme}
\label{sec:V1D}

We use our modified V1D version \citep{KK2019} that is compatible with the input physics of Appendix~\ref{sec:input}. Our isotope list is the $69$-isotope list calibrated by \citet{Kushnir2020}, which allows the calculation of $\nimass$ with an accuracy greater than one percent. We safely ignore weak reactions and thermal neutrino emissions \citep{Kushnir2020}. We do not use linear artificial viscosity, the Courant time-step factor is $0.25$, and the maximum relative change of the density in each cell during a time-step is set to $0.01$. 

We use our recently developed, accurate and efficient burning scheme that allows the structure of TNDW to be resolved \citep{KK2019}. The numerical scheme contains a burning limiter that broadens the width of the TNDW while accurately preserving its internal structure. The burning limiter limits the changes in both energy and composition to a fraction $f$ during cell sound crossing time (for faster changes, all rates are normalized by a constant factor to limit the changes). Burning is calculated in-situ by employing the required large-networks without the use of post-processing or pre-describing the conditions behind the TNDW. The scheme was tested against accurate solutions of the structure of TNDW with resolutions that are typical for multidimensional (multi-D) full-star simulations, yielding an accuracy that is better than one percent for the resolved scales (where the burning limiter is not applied) and a few percent for unresolved scales (broadened by the burning limiter). Burning is not allowed on shocks (identified as cells where $q_{v}/p>0.1$, where $q_{v}$ is the artificial viscosity and $p$ is the pressure). The allowed error tolerance for the burning integration is $\delta_{B}=10^{-8}$ \citep[see][for details]{KK2019}, and we use $f=0.1$ for the burning limiter. 

The mesh includes only the WD, with the outer numerical node at $r_p$. The inner boundary condition is of a solid wall and the outer boundary velocity is the assumed piston velocity. All cells are initially of an equal size, $\Delta x_{0}$, and the density in each cell is determined by interpolation from the original WD profile to the center of the cell.

Since the initial profile is interpolated to the mesh, it is not in strict hydrostatic equilibrium. We therefore only activate cells that are just in front of the leading shock. This is done by finding the innermost active cell with $q_{v}/p>10^{-3}$, and then activating its inner node. Initially, all cells within $[r_{p}-5\Delta x_{0},r_{p}]$ are activated. The simulations are stopped $0.25\,\textrm{s}$ after the imploding shock focuses to the center. This is after the outgoing shock (or a TNDW in the case of core ignition) had travelled a significant fraction of the WD radius, but before reaching the outer node. 


\section{The work delivered by the Helium shell}
\label{sec:work}

\begin{figure}
\includegraphics[width=0.48\textwidth]{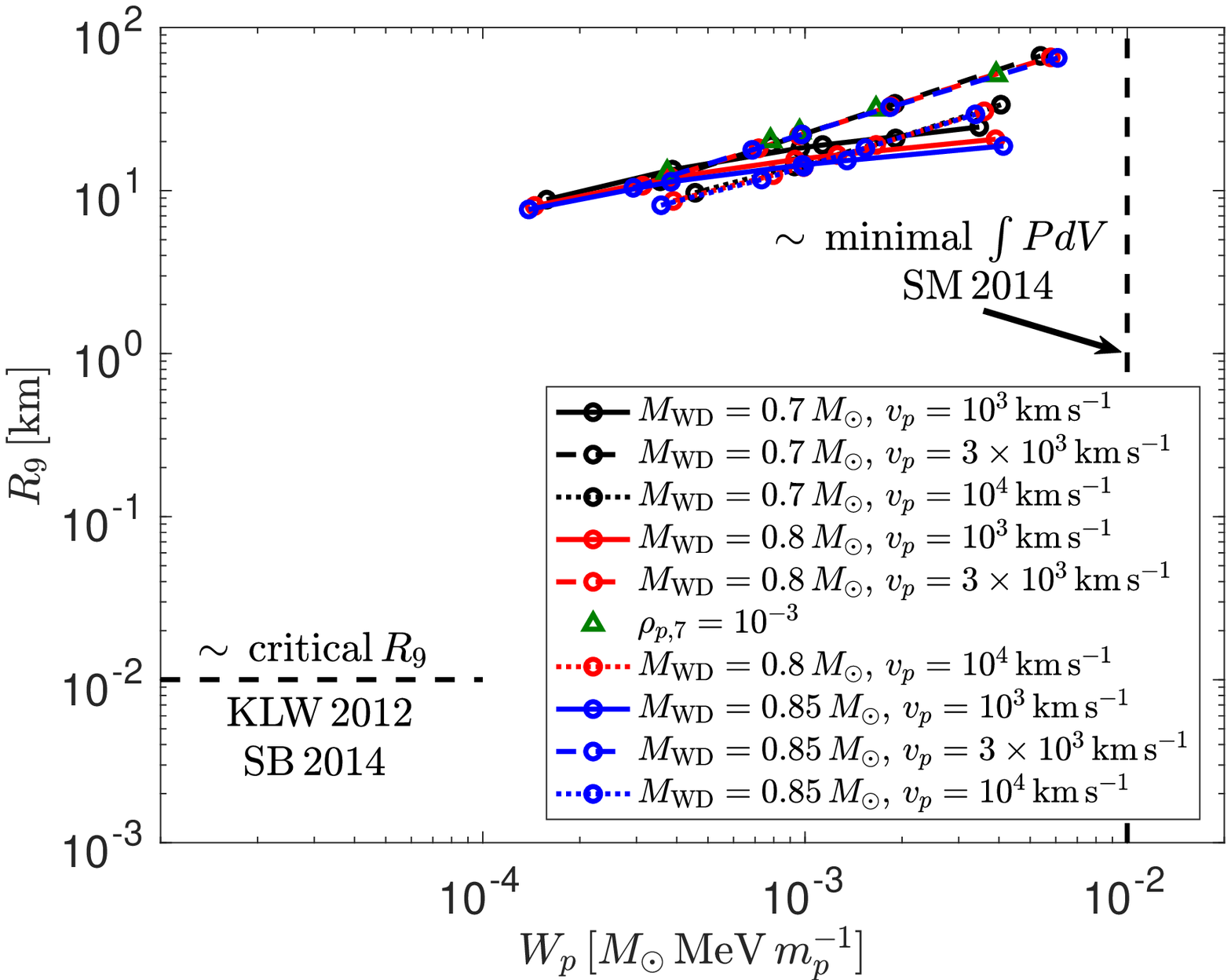}
\caption{The converged values of $R_{9}$ (the radius in which the shock velocity is $10^{9}\,\textrm{cm}\,\textrm{s}^{-1}$) as a function of the work delivered by the piston, $W_p$. We present results for  $\MWD=0.7,0.8,0.85\,M_{\odot}$ (black, red and blue, respectively) and $v_{p}=1,3,10\times10^{3}\,\textrm{km}\,\textrm{s}^{-1}$ (solid, dashed and dotted, respectively). For each combination of $\MWD$ and $v_p$, we choose a few values of $t_p$ whose $W_p$ values bracket $10^{-3}\,M_{\odot}\,\textrm{MeV}\,m_p^{-1}$. For $W_p=10^{-3}\,M_{\odot}\,\textrm{MeV}\,m_p^{-1}$, we get $R_9\sim10-20\,\textrm{km}$, with a weak dependence on the value of $v_p$. The obtained $R_9$ are larger by orders of magnitude than the critical $R_9$ for ignition \citep[$\mysim0.01\,\textrm{km}$;][]{KLW2012,Shen2014}. This result by itself is already sufficient to show that ignition in these cases is robust. A $W_p=10^{-3}\,M_{\odot}\,\textrm{MeV}\,m_p^{-1}$ is well below the minimal work that can be delivered by a TNDW that propagates within the helium shell \citep[$\mysim10^{-2}\,M_{\odot}\,\textrm{MeV}\,m_p^{-1}$][]{ShenMoore2014}. Simulations with $\MWD=0.8\,M_{\odot}$, $v_{p}=3\times10^{3}\,\textrm{km}\,\textrm{s}^{-1}$ and $\rho_{p,7}=10^{-3}$ (green triangles) show that the value of $\rho_p$ has but a small effect on our results. 
\label{fig:R9_PdV}}
\end{figure}

We first perform calculations without nuclear burning for $\MWD=0.7,0.8,0.85\,M_{\odot}$ and $v_{p}=1,3,10\times10^{3}\,\textrm{km}\,\textrm{s}^{-1}$. For each combination of $\MWD$ and $v_p$, we choose a few values of $t_p$ whose corresponding $W_p$ values bracket $10^{-3}\,M_{\odot}\,\textrm{MeV}\,m_p^{-1}$, and we determine $R_9$ in each case. Since direct differencing of the shock position is too noisy to allow the determination of the shock velocity, we instead fit the shock position to a power law, $R_s=A(t-t_0)^\alpha$, which is the exact self-similar solution for an ideal gas and roughly holds in our case (with $\alpha\approx0.75$). We determine $R_9$ from the fit parameters. The fit is performed over times where $R_s$ is in the range $[10,100]\Delta x_0$. We increase the resolution in each case, typically up to $\Delta x_0\lesssim1\,\textrm{km}$, until $R_9$ converges to a few percent (and $R_9$ is within the range of the fit). The converged values of $R_9$ are presented in Figure~\ref{fig:R9_PdV}. As can be seen in the figure, for $W_p=10^{-3}\,M_{\odot}\,\textrm{MeV}\,m_p^{-1}$, we get $R_9\sim10-20\,\textrm{km}$, with a weak dependence on the value of $v_p$. The obtained $R_9$ are larger by orders of magnitude than the critical $R_9$ for ignition ($\mysim0.01\,\textrm{km}$). This result by itself is already sufficient to show that ignition in these cases is robust\footnote{We verified that this is the case also for $\MWD=0.5\,M_{\odot}$, the lowest mass CO WDs, with $R_9\gtrsim10\,\rm{km}$.}. Nevertheless, we want to verify that ignition is indeed obtained. As we show in Section~\ref{sec:example}, the size of the hot spot is $\mysim R_9/30$, which allows us to resolve the hot spot with modest resolution. A higher resolution would be required for lower values of $W_p$, but since $W_p=10^{-3}\,M_{\odot}\,\textrm{MeV}\,m_p^{-1}$ is already well below the minimal work that is delivered by a TNDW that propagates within the helium shell \citep[$\mysim10^{-2}\,M_{\odot}\,\textrm{MeV}\,m_p^{-1}$][]{ShenMoore2014}, there is no need to demonstrate ignition for lower values of $W_p$. Simulations for $\MWD=0.8\,M_{\odot}$, $v_{p}=3\times10^{3}\,\textrm{km}\,\textrm{s}^{-1}$ and $\rho_{p,7}=10^{-3}$ (green triangles in the figure) produced only slightly modified results. 

The reason that the shock dynamics mostly depends on $W_p$ and not on the exact details of the piston stems from the dynamics of imploding shocks. As the shock accelerates, a sonic point forms behind the shock that separates the shock from the boundary. The flow between the shock and the sonic point approaches a self-similar flow \citep{Guderley42}, which is independent of the boundary conditions. Therefore, the piston can only affect the shock before the sonic point forms, and the flow between the shock and the sonic point mostly depends on the work delivered to the plasma prior to this time, as we demonstrated above.  

We next verify directly that indeed $W_p=10^{-3}\,M_{\odot}\,\textrm{MeV}\,m_p^{-1}$ is sufficient for core ignition. We calculate the core ignition with nuclear burning for each combination of $\MWD$ and $v_p$, with $t_p$ chosen so as to deliver exactly $W_p=10^{-3}\,M_{\odot}\,\textrm{MeV}\,m_p^{-1}$ ($9$ combinations and one more with $\rho_{p,7}=10^{-3}$ from above, see Table~\ref{tbl:simulations}). In all cases, ignition is obtained once a high-enough resolution is employed, typically $\Delta x_0\lesssim1\,\textrm{km}$, and the hot spot is resolved (see example in Section~\ref{sec:example}). Following ignition, two TNDWs are formed, one that propagates inwards and one that propagates outwards, following which the star explodes. We, therefore, prove that $W_p=10^{-3}\,M_{\odot}\,\textrm{MeV}\,m_p^{-1}$ is sufficient to ignite TNDW for all considered cases. 

\begin{table*}
\caption{The configurations calculated with nuclear burning. The calibrated $t_p$ values  to deliver exactly $W_p=10^{-3}\,M_{\odot}\,\textrm{MeV}\,m_p^{-1}$ are given in the 4th column, as a function of $\MWD$ (1st column), $\rho_{p,7}$ (2nd column) and $v_p$ (3rd column).}
\begin{tabular}{|c||c||c||c|}
\hline
$\MWD\,[M_{\odot}]$  & $\rho_{p,7}$ &$v_p\,[10^3\,\textrm{km}\,\textrm{s}^{-1}]$    & $t_p\,[\textrm{s}]$ \\ \hline
0.7	&  0.01 & 1 & 1.82  \\
0.7	&  0.01 &   3 & 0.24  \\
0.7	&  0.01 &   10 & 0.0107 \\
0.8	&  0.01 &   1 & 1.7 \\
0.8	&  0.01 &   3 & 0.25 \\
0.8	&  $10^{-3}$ &   3 & 0.55 \\
0.8	&  0.01 &   10 & 0.0123 \\
0.85	&  0.01 &   1 & 1.7 \\
0.85	&  0.01 &   3 & 0.26 \\
0.85	&  0.01 &   10 & 0.0133 \\
\hline
\end{tabular}
\centering
\label{tbl:simulations}
\end{table*}


\section{An example of a resolved ignition}
\label{sec:example}

In this section, we provide an example of a successful core ignition with $\MWD=0.8\,M_{\odot}$, $v_p=3\times10^{3}\,\textrm{km}\,\textrm{s}^{-1}$ and $t_p=0.25\,\textrm{s}$ (leading to $W_p\approx10^{-3}\,M_{\odot}\,\textrm{MeV}\,m_p^{-1}$). In this case, we find $R_9\approx20\,\textrm{km}$ by calculating without nuclear burning. The required resolution for resolving the hot spot in a calculation that includes nuclear burning is determined by the size of the hot spot, $\Delta r$. We first provide a general analysis of the size of the hot spot, and then we present the numerical results from the example calculation. 

Consider the shock wave as it approaches $R_9$. We can assign for each mass coordinate, $m$, an explosion time, $t_{\rm{exp}}(m)$, which corresponds to the time in which burning becomes very fast (we set $t=0$ as the time in which the shock reaches the center). We can therefore write
\begin{eqnarray}\label{eq:tsh}
t_{\rm{exp}}(m)=t_{\rm{sh}}(m)+t_{\varepsilon}\left[T_{\rm{sh}}\left(t_{sh}(m)\right)\right],
\end{eqnarray}
where $t_{\rm{sh}}(m)$ is the time in which the shock reaches the mass element (and it is negative with our choice of $t=0$) and $t_{\varepsilon}$ is the induction time, which depends on the post-shock temperature, $T_{\rm{sh}}$. Since the induction time decreases with increasing $T_{\rm{sh}}$, which is higher as the shock accelerates, the function $t_{\rm{exp}}(m)$ has a minimum at some distance behind the shock. The induction time at this position can be found by a direct differentiation of Equation~\eqref{eq:tsh}:
\begin{eqnarray}\label{eq:der1}
0=\frac{dt_{\rm{exp}}}{dt_{\rm{sh}}}=1+\frac{dt_{\varepsilon}}{dT_{\rm{sh}}}\frac{dT_{\rm{sh}}}{dt_{\rm{sh}}}=1-\nu\beta\frac{t_{\varepsilon}}{t_{\rm{sh}}},
\end{eqnarray}
where we assume $t_{\varepsilon}\propto T_{\rm{sh}}^{-\nu}$ and we define
\begin{eqnarray}\label{eq:beta}
\frac{dT_{\rm{sh}}}{dt_{\rm{sh}}}=\beta\frac{T_{\rm{sh}}}{t_{\rm{sh}}}.
\end{eqnarray}
From the cross-section of $^{12}$C$+^{12}$C, we find for $T_{\rm{sh},9}\approx2-4$ that $\nu\approx20-15$. We can estimate $\beta$ from the self-similar solution for the shock radius
\begin{eqnarray}\label{eq:beta}
R_{\rm{sh}}\propto t_{\rm{sh}}^{\alpha}\Rightarrow T_{\rm{sh}}\propto\dot{R}^{2}_{\rm{sh}}\propto t_{\rm{sh}}^{2(\alpha-1)}\Rightarrow \beta=2(\alpha-1).
\end{eqnarray}
We found in Section~\ref{sec:work} for all cases $\alpha\approx0.75$, such that $\beta\approx-0.5$, and from Equation~\eqref{eq:der1}, we get
\begin{eqnarray}\label{eq:der2}
t_{\varepsilon}=\frac{t_{\rm{sh}}}{\nu\beta}\approx-\frac{t_{\rm{sh}}}{10}. 
\end{eqnarray}
For a successful ignition, the energy release time scale, $\varepsilon/\dot{q}\sim t_{\varepsilon}$, where $\dot{q}$ is the energy injection from burning and $\varepsilon$ is the thermal energy, needs to be shorter than the sound-crossing time of the hot spot, $\Delta r/c_s$, where $c_s$ is the speed of sound \citep{zeldovich,Kushnir2015}. Therefore, the size of the hot spot during ignition satisfies $\Delta r\approx c_s t_{\varepsilon}\approx-c_s t_{\rm{sh}}/10\approx R_{\rm{sh}}/30$, where we use $c_s\approx-0.4\dot{R}_{\rm{sh}}\approx0.4\alpha R_{\rm{sh}}/t_{\rm{sh}}$\footnote{For an ideal gas, the exact expression is $c_s=-\sqrt{2\gamma(\gamma-1)/(\gamma+1)^2}\dot{R}_{\rm{sh}}$, where $\gamma$ is the adiabatic index of the gas. We use $\gamma=4/3$ for our estimate \citep[that corresponds to $\alpha\approx0.73$, see, e.g.,][]{KLW2012}.}. 

\begin{figure}
\includegraphics[width=0.48\textwidth]{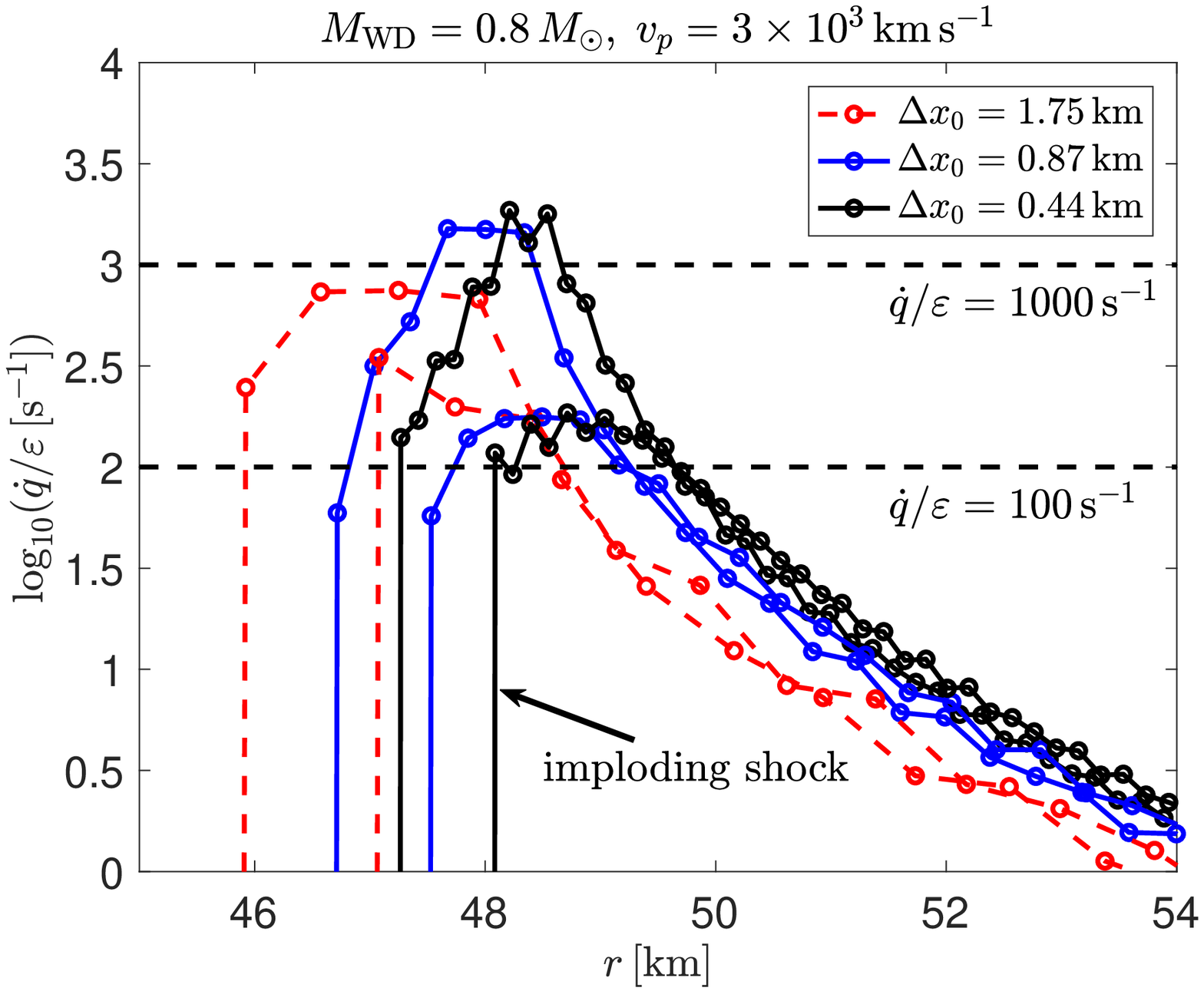}
\caption{A resolved ignition in the $\MWD=0.8\,M_{\odot}$, $v_p=3\times10^{3}\,\textrm{km}\,\textrm{s}^{-1}$ and $t_p=0.25\,\textrm{s}$ (leading to $W_p\approx10^{-3}\,M_{\odot}\,\textrm{MeV}\,m_p^{-1}$) case. Snapshots of the relative burning rate, $\dot{q}/\varepsilon$, are presented in a series of simulations with increasing resolutions ($\Delta x_0=1.75,0.87,0.44\,\textrm{km}$ in red, blue and black, respectively), at two times separated by about $0.1$ ms around the time of ignition. Note that the plasma at the ignition location is compressed by a factor of $\mysim3$ relative to the initial density. 
\label{fig:M08_vp3e8_ignition}}
\end{figure}

We demonstrate the ability to resolve the ignition region in Figure~\ref{fig:M08_vp3e8_ignition}. The burning rate $\dot{q}/\varepsilon$ is shown in the vicinity of the ignition location, at two snapshots separated by about $0.1\,\textrm{ms}$ around the onset of ignition. The burning runaway is obtained at $R_{\rm{sh}}\approx50\,\textrm{km}$ (with $\dot{R}_{\rm{sh}}\approx7.5\times10^{3}\,\textrm{km}\,\textrm{s}^{-1}$), which is larger by a factor of $\approx2.5$ from the rough estimate of $R_9$. Using our estimate from above, we expect a hot spot size of $\Delta r\approx1.5\,\textrm{km}$. As can be seen in the figure, by the time of the second snapshot, a region with such a width indeed forms, and our level of resolution is sufficient to resolve it.  The speed of sound in this region is $c_s\approx 4.5\times 10^{3} \,\textrm{km}\,\textrm{s}^{-1}$ (which roughly agrees with our estimate from above) with a sound-crossing time of $\approx0.3\,\textrm{ms}$. The hot spot is producing energy at a rate above $100\,\textrm{s}^{-1}$, which more than doubles within $0.1\,\textrm{ms}$. A significant amount of energy is released within $\mysim0.1\,\textrm{ms}$ and the sound waves do not have sufficient time to distribute the excess pressure, resulting in the formation of two detonation fronts a short time later (not shown here). 

\section{Summary and discussion}
\label{sec:discussion}

We used 1D numerical simulations (described in Section~\ref{sec:setup}) to show that if a TNDW is able to propagate within the helium shell, then ignition within the CO core is guaranteed (Section~\ref{sec:work}). We demonstrated this for $\MWD=0.7,0.8,0.85\,M_{\odot}$, where we were able to numerically resolve the ignition in a full-star simulation for the first time. Previous calculations of the DDM \citep[see, e.g.,][for recent studies]{Polin2019,Townsley2019,Boos2021} used an unstable burning scheme (or implemented a stabilizing burning limiter, but relaxed it during ignition phases), in which the ignition is presumably a result of numerical instability. For example, a "scissors mechanism" was claimed for the CO core, opposite to helium detonation ignition spot \citep{Gronow2020,Gronow2021a,Gronow2021b}. In these simulations, the ignition was obtained in a density of $\rho_7\approx0.3$ and a temperature of $T_9\approx2.4$, where the burning time is $\approx4\times10^{-5}\,\textrm{s}$ (for the relevant composition of $90\%$ $^{4}$He, $5\%$ $^{12}$C, and $5\%$ $^{16}$O by mass), which is much shorter than the sound-crossing time of the numerical cells, $\approx4\times10^{-3}\,\textrm{s}$ (for the implemented $20\,\textrm{km}$ resolution). Ignition under these conditions is obtained in the regime of numerical instability \citep{Kushnir2013}.  

Our findings do not support the possibility that in the DDM the ignition of low-mass CO cores is suppressed and that the probability of low-mass WDs to explode is much lower than the probability of high-mass WDs to explode. This is in apparent contradiction with the observations that, on the one hand, the rate of low-luminosity ($\nimass\approx0.15\,M_{\odot}$) SNe Ia is suppressed by a factor of $\mysim10$ compared to the more common $\nimass\approx0.7\,M_{\odot}$ events \citep{Sharon2022}, and that, on the other hand, $\approx0.85\,M_{\odot}$ WDs (the progenitors of $\nimass\approx0.15\,M_{\odot}$ in the DDM) are much more common (by a factor of $\mysim10$) than $\approx1.05\,M_{\odot}$ WDs (the progenitors of $\nimass\approx0.7\,M_{\odot}$ in the DDM) \citep{Bergeron2019}; see Figure~\ref{fig:rate}. Out results indicate, therefore, that the DDM provides no natural explanation for the low-luminosity suppression of SNe Ia. 

One caveat to the arguments presented here is that perhaps the probability of a WD to be involved in a binary that leads to the required conditions for the DDM to operate is much lower for low-mass WDs than for high-mass WDs (see Section~\ref{sec:Introduction}). Additional observational and theoretical work is required to measure the mass distribution of the primary WDs in close binaries. It is also possible that the ignition probability of a TNDW within the Helium shell is suppressed for low-mass WDs (see Section~\ref{sec:Introduction}). A study of this possibility is outside the scope of this work.

Another caveat is that we used a 1D approximation to describe the work delivered by the helium shell to the CO core. This approximation cannot take into account the multi-D structure of the imploding shock. For example, the finite travel time of the TNDW in the helium shell around the WD should have a stronger effect on lower-mass WDs due to their larger radii, which may result a larger asphericity of the imploding shock and a weakening of its strength. This effect may also lead to a shift of the focal point to lower densities away from the center of the WD. While the effect of this perturbation does not seem to be large \citep[see, e.g.,][]{Townsley2019,Boos2021} and the obtained $R_9$ is orders of magnitude greater than the critical $R_9$ (even at the low densities of  $\MWD=0.5\,M_{\odot}$, see Section~\ref{sec:work}) multi-D simulations are required to verify the robustness of the ignition. Nevertheless, since WDs in the mass range of $\MWD=0.85-1.05\,M_{\odot}$ have a similar structure, we do not expect a strong dependence of the ignition properties on the WD mass in this range. 
 
The DDM's inability to explain the low-luminosity suppression joins other challenges this model faces. Recently, \citet{Kushnir2020} used the observed $t_0-\nimass$ relation \citep{Sharon2020a}, where $t_0$ is the $\gamma$-rays escape time from the ejecta (measured to a few percent accuracy), to reveal a clear tension between the predictions of SCD and the observed positive correlation between $t_0$ and $\nimass$. SCD predicts an anti-correlation between $t_0$ and $\nimass$, with $t_0\approx30\,\textrm{day}$ for luminous ($\nimass\gtrsim0.5\,M_{\odot}$) SNe Ia, while the observed $t_0$ is in the range of $35-45\,\textrm{day}$. They showed that various uncertainties related to the physical processes and to the initial profiles of the WD are unlikely to resolve the tension with observations, although they can deteriorate the agreement with observations of low-luminosity SNe Ia. While there are some reasonable initial compositions and reaction rate values for which DDM successfully explains the low-luminosity part of the $t_0-\nimass$ relation, this model may be in conflict with the observed $^{56}$Ni mass-weighted line-of-sight velocity distribution for a large fraction of these events, as measured from nebular spectra  \citep{Dong2015,Dong2018,Vallely2020}. Specifically, the $^{56}$Ni velocity distribution is either double-peaked or highly shifted, which so far has not been predicted by the DDM. 

The low-luminosity suppression poses a challenge for any model. As far as we know, such a suppression is not predicted by Chandrasekhar-mass models, and anyway, they are unable to explain the $t_0-\nimass$ relation for low-luminosity SNe Ia \citep{Wygoda2019,Sharon2020b}. Perhaps the direct-collision model \citep{Kushnir2013} can explain this type of suppression, if the ignition depends strongly on the impact parameter of the colliding WDs. Such a phenomena can only be studied with 3D simulations.

\section{Data availability}

The data underlying this article will be shared following a reasonable request to the corresponding author.

\section*{Acknowledgements}
We thank Boaz Katz for suggesting the topic of this paper and for contributing to the analytical derivation of the hot spot size. We thank Subo Dong for useful discussions. DK is supported by the Israel Atomic Energy Commission -- The Council for Higher Education -- Pazi Foundation -- a research grant from The Abramson Family Center for Young Scientists -- and an ISF grant.  






\begin{appendix}

\section{Input physics}
\label{sec:input}

Our input physics, which we briefly summarize below, are the ones used by \citet{KK2019}. A detailed description can be found in \citep{Kushnir2019}.

The nuclear masses are taken from the file \textsc{winvn\_v2.0.dat}, which is available through the JINA reaclib data base\footnote{http://jinaweb.org/reaclib/db/} \citep[JINA,][]{Cyburt2010}. For the partition functions, $w_{i}(T)$, we used the fit of \citet{Kushnir2019} for the values that are provided in the file \textsc{winvn\_v2.0.dat} over some specified temperature grid. The forward reaction rates are taken from JINA (the default library of 2017 October 20). All strong reactions that connect between isotopes from the list are included. Inverse reaction rates were determined according to a detailed balance. Enhancement of the reaction rates due to screening corrections is described at the end of this section. We further normalized all the channels of the $^{12}$C+$^{16}$O and $^{16}$O+$^{16}$O reactions such that the total cross-sections are identical to the ones provided by \citet{CF88}, while keeping the branching ratios provided by JINA. We ignore weak reactions and thermal neutrino emissions. 

The EOS is composed of contributions from electron--positron plasma, radiation, ideal gas for the nuclei, ion--ion Coulomb corrections and nuclear level excitations. We use the EOS provided by {\sc MESA} for the electron--positron plasma, for the ideal gas part of the nuclei, for the radiation and for the Coulomb corrections (but based on \citet{Chabrier1998} and not on \citet{Yakovlev1989}, see below). The electron--positron part is based on the \textit{Helmholtz} EOS \citep{Timmes00}, which is a table interpolation of the Helmholtz free energy as calculated by the Timmes EOS\footnote{http://cococubed.asu.edu/} \citep{Timmes1999} over a density-temperature grid with $20$ points per decade. This is different from \citet{Kushnir2019}, where the Timmes EOS was used for the electron--positron plasma, since the \textit{Helmholtz} EOS is more efficient and because the internal inconsistency of the \textit{Helmholtz} EOS \citep[see][for details]{Kushnir2019} is small enough within the regions of the parameter space studied here. We further include the nuclear level excitation energy of the ions, by using the $w_{i}(T)$ from above.

We assume that the Coulomb correction to the chemical potential of each ion is given by $\mu_{i}^{C}=k_{B}Tf(\Gamma_{i})$ and is independent of the other ions \citep[linear mixing rule (LMR),][]{Hansen1977}, where $k_{B}$ is Boltzmann's constant, $\Gamma_{i}=Z_{i}^{5/3}\Gamma_{e}$ is the ion coupling parameter, where $Z_i$ is the proton number, and $\Gamma_{e}\approx(4\upi\rho N_{A} Y_{e}/3)^{1/3}e^{2}/k_{B}T$ is the electron coupling parameter, where $N_{A}$ is Avogadro's number and $Y_e\approx\sum_i X_i Z_i/A_i$ is the electron fraction. We use the three-parameter fit of \citet{Chabrier1998} for $f(\Gamma)$.  Following \citet[][]{Khokhlov88}, we approximate the LMR correction to the EOS by $f(\Gamma)$ for a `mean' nucleus $\Gamma=\bar{Z}^{5/3}\Gamma_{e}$, where
\begin{eqnarray}
\bar{Z}=\frac{\sum_i Y_i Z_i}{\sum_i Y_i}.
\end{eqnarray}
The screening factor for a thermonuclear reaction with reactants $i=1,..,N$ and charges $Z_{i}$ is determined from detailed balance \citep{KushnirScreen}:
\begin{eqnarray}\label{eq:NSE screening}
\exp\left(\frac{\sum_{i=1}^{N}\mu_{i}^{C}-\mu_{j}^{C}}{k_{B}T}\right),
\end{eqnarray}
where isotope $j$ has a charge $Z_{j}=\sum_{i=1}^{N}Z_{i}$ \citep[same as equation~(15) in][for the case of $N=2$]{Dewitt1973}.

\end{appendix}

\bsp	
\label{lastpage}
\end{document}